\documentclass[9pt,journal]{IEEEtran} \newcommand{\figsizefull}{0.99}

\usepackage[cmex10]{amsmath}
\usepackage{amsfonts}
\usepackage{amssymb}
\usepackage{amsopn}
\usepackage{graphicx}
\usepackage{subfigure}
\usepackage{epsfig}
\usepackage{color}

\newtheorem{prop:limit}{Proposition}
\newtheorem{prop:unit}{Lemma}
\newtheorem{prop:threshold}{Theorem}

\newtheorem{def:admissible}{Definition}
\newtheorem{def:antisymmetric}[def:admissible]{Definition}
\newtheorem{def:dominant}[def:admissible]{Definition}
\newtheorem{def:unimodal}[def:admissible]{Definition}

\DeclareMathOperator*{\minimize}{minimize}
\newcommand{\ud}{\,\mathrm{d}}

\newcommand{\bo}[1]{\boldsymbol{#1} }
\makeatletter
\def\blfootnote{\xdef\@thefnmark{}\@footnotetext}
\makeatother

\begin{document}

\markboth{TO APPEAR in IEEE TRANSACTIONS ON SIGNAL PROCESSING}{Kar \MakeLowercase{\textit{et al.}}: Optimal Identical Binary Quantizer Design for Distributed Estimation}

\title{Optimal Identical Binary Quantizer Design for Distributed Estimation}

\author{Swarnendu~Kar,~\IEEEmembership{Student~Member,~IEEE,}
        Hao~Chen,~\IEEEmembership{Member,~IEEE,}
        and~Pramod~K.~Varshney,~\IEEEmembership{Fellow,~IEEE}
}

\maketitle

\begin{abstract}
We consider the design of identical one-bit probabilistic quantizers for distributed estimation in sensor networks. We assume the parameter-range to be finite and known and use the maximum Cram\'er-Rao Lower Bound (CRB) over the parameter-range as our performance metric. We restrict our theoretical analysis to the class of antisymmetric quantizers and determine a set of conditions for which the probabilistic quantizer function is greatly simplified. We identify a broad class of noise distributions, which includes Gaussian noise in the low-SNR regime, for which the often used threshold-quantizer is found to be minimax-optimal. Aided with theoretical results, we formulate an optimization problem to obtain the optimum minimax-CRB quantizer. For a wide range of noise distributions, we demonstrate the superior performance of the new quantizer - particularly in the moderate to high-SNR regime.
\end{abstract}

\begin{IEEEkeywords}
Minimax CRLB, dithering, probabilistic quantization, distributed estimation
\end{IEEEkeywords}

\section{Introduction}
Distributed estimation is a classical problem in statistical signal processing, where a fusion center (FC) receives compressed data from several information nodes and infers the parameter(s) of the underlying process. Consider a distributed estimation problem with $N$ sensors collecting noisy observations of an unknown but fixed scalar parameter $\theta$ such that the local sensor observations $\bo X = [X_1,X_2,\ldots X_N]'$ are independent and identically distributed (i.i.d.), i.e., $f(\bo X,\theta)=\prod_{n=1}^N f(X_n,\theta )$, where $f(\bo X,\theta )$ and $f(X_n,\theta )$ are known probability density functions (pdf). One example of such a model is the location estimation problem with additive noise,
\begin{align}
    X_n=\theta +W_n, \qquad 1\le n \le N, \label{def:X}
\end{align}
where the noise samples $\bo W = [W_1,W_2,\ldots W_N]'$ are zero-mean, additive, independent, and identically distributed with symmetric pdf $f(w)$ and variance $\sigma^2$. In many practical applications, the dynamic range of $\theta$ is often assumed to be known, such that $\theta \in [\theta_0-\Delta,\theta_0+\Delta]$ where $\theta_0$ and $\Delta$ are known constants. Without loss of generality, we assume $\theta_0=0$ and $\Delta=1$ and confine our attention to $\theta\in[-1,1]$ in the rest of this paper. \blfootnote{S. Kar and P. K. Varshney are with the Department of Electrical Engineering and Computer Science, Syracuse University, Syracuse, NY, 13244 USA, e-mail: \{swkar,varshney\}@syr.edu. H. Chen is with College of Engineering, Boise State University, Boise, ID 83725. e-mail: \{haochen\}@boisestate.edu.} \blfootnote{This research was partially supported by the National Science Foundation under Grant No. $0925854$ and the Air Force Office of Scientific Research under Grant No. FA-9550-10-C-0179. } \blfootnote{Copyright (c) 2012 IEEE. Personal use of this material is permitted. However, permission to use this material for any other purposes must be obtained from the IEEE by sending a request to pubs-permissions@ieee.org.}

As an application instance of this problem domain, one can consider an environmental monitoring system consisting of a central base station communicating with multiple thermal sensors with limited energy deployed over a region in a dense manner, so that they are more-or-less recording the same temperature at any given time. The redundancy in the number of sensors serves to increase robustness of the network, share the power resources and increase the lifetime of the monitoring system. The objective of the sensor network is to monitor the temperature in the region throughout the day, though the diurnal temperature variation (say $10-40^{\circ}$C) is roughly known.

\subsection{Identical one-bit quantizers}
Since the channel capacity of links between sensors and the fusion center and the energy resources for transmission in the battery-powered sensor nodes can be severely limited, we assume that each sensor performs a binary quantization and transmits only one-bit of information to the FC. With an appeal to symmetry, each sensor is designed to employ an identical quantization rule.

A one-bit quantizer can be defined as a mapping from the observation space $\mathbb R$ to a symbol set of size $2$, say $\{\mathcal S_0,\mathcal S_1\}$. Such a mapping can be expressed in two forms. In the often used \emph{function-form} description, a quantizer explicitly maps its input $X_n$ to the output $Y_n$ through a function $\varphi: \mathbb R\rightarrow \{\mathcal S_0,\mathcal S_1\}$. For example, the function-form description of a zero-threshold quantizer is
\begin{align}
Y_n=\varphi_T (X_n) \triangleq \left\{
\begin{array}{cc}
\mathcal S_1, & X_n \geq 0 \\
\mathcal S_0, & X_n < 0
\end{array}  \right., \forall n. \label{qtzr:th:fnc}
\end{align}
Alternatively, in the \emph{probability-form} description, a quantizer is defined as the conditional probability $\gamma: \mathbb R\rightarrow [0,1]$  of the output being a particular symbol (say $\mathcal S_1$) given an input $X_n$,
\begin{align}
\gamma(X_n)\triangleq P(Y_n=\mathcal S_1|X_n), \forall n. \label{def:gamma}
\end{align}
For example, the equivalent probability-form description of the zero-threshold quantizer \eqref{qtzr:th:fnc} is
\begin{align}
\gamma_T (X_n) \triangleq \left\{
\begin{array}{cc}
1, & X_n \geq 0 \\
0, & X_n < 0
\end{array}  \right., \forall n, \label{qtzr:th:prob}
\end{align}
which we would refer to as the \emph{Threshold Quantizer}. In this paper, we will use the probability-form description \eqref{def:gamma} for analysis and subsequent design of quantizers. Here by allowing $\gamma $ to take any value between $0$ and $1$, we consider all possible local quantization rules \cite{Chen10}, i.e., the quantization rule can be either deterministic (e.g., single threshold quantizer \cite{Rib06},\cite{Venkita07}) or probabilistic (e.g., dithered quantizer, i.e., some noise added to the signal before quantization \cite{Gray93},\cite{Papado01}).

\subsection{Reliable Transmission}
In this paper, we assume that the stringent source-rate constraint (1-bit per observation) frees up resources so that adequate channel-coding is undertaken to counter noise/fading phenomena in the communication channel. As a result, the compressed information in $\bo Y=[Y_1,Y_2,\ldots,Y_N]$ is assumed to be obtained in a lossless fashion at the FC, an assumption that is consistent with several previous research contributions on this topic \cite{Rib06,Luo05}. However, in the more general scenario, absence of sufficient resources for adequate channel-coding may result in lossy transmission of $\bo Y$ - an issue also considered by several researchers \cite{Aysal08,Wu09}. Though the results in this paper can be extended to noisy channel scenario, we shall not discuss this extension here due to space constraints.

\subsection{Performance metric}
The goal of the fusion center is to use the quantized observations and obtain an estimate of the location parameter $\widehat{\theta}$ using an estimator $h(\bo Y,\gamma)$. The problem setup is summarized by the Markov chain,
\begin{align}
    \theta \stackrel{f(w)}{\rightarrow} \bo X \stackrel{\gamma}{\rightarrow} \bo Y \stackrel{h}{\rightarrow} \widehat{\theta}.
\end{align}
Let $\widehat{\theta}$ be an unbiased estimator of $\theta$. It is well known that the variance of any unbiased estimator is lower bounded by the Crame\'r-Rao lower bound  ($\text{CRB}$) and that the CRB is asymptotically achieved by using the Maximum-Likelihood (ML) estimator (see \cite{Kay93}). Let $g(\theta)$ denote the probability that the quantizer output is $\mathcal S_1$ when the original parameter is $\theta$,
\begin{align}
g(\theta)&=P(Y_n=\mathcal S_1|\theta) = \mathbb{E}_{W_n}(\gamma(\theta+W_n)) \nonumber \\
&=\int_{-\infty}^{\infty}{\gamma(x)f(x-\theta) \ud x}. \label{def:g}
\end{align}
Then the sample mean $\overline{\bo Y}=\frac{1}{N}\sum_{i=1}^N Y_i$ is the ML-estimate of $g(\theta)$ and from the functional invariance property, we have
\begin{align}
\widehat{\theta}_{\text{ML}}=g^{-1}(\overline{\bo Y}). \label{theta:ml}
\end{align}
For $N$ independent observations, the variance of $\widehat{\theta}_{\text{ML}}$ satisfies
\begin{align}
\mathbb{E}\{(\theta-\widehat{\theta})^2\} &\geq \frac{1}{N}\frac{1}{I(\theta)} \triangleq   \frac{1}{N} \text{CRB}(\theta,\gamma,f),
\end{align}
where $I(\theta)=-\mathbb{E}_W\left[ \nabla^2_\theta \ln p(Y_1;\theta) \right]$ is the Fisher Information (FI) for one sensor output and the equality can be achieved asymptotically \cite{Venkita07}. In case of binary quantization, FI can be expressed as (see \cite{Chen10}),
\begin{align}
I(\theta)=\frac{\left(g'(\theta)\right)^2}{g(\theta)(1-g(\theta))}. \label{Itheta:g:g1}
\end{align}

In general, $\text{CRB}(\theta,\gamma;f)$ is a function of the unknown parameter $\theta$, i.e., the quantizer $\gamma$ may result in a high CRB for one $\theta$ and a low CRB for another. To ensure accurate estimation over the entire parameter range, we use the maximum possible estimation variance or the \emph{maximum-CRB}
\begin{align}
\phi(\gamma,f)=\sup_{\theta \in (-1,1)} \text{CRB}(\theta,\gamma,f), \label{def:phi:gamma}
\end{align}
as our performance metric.

Although it is relatively easy to obtain $\phi(\gamma,f)$ for a given noise probability distribution function $f(.)$ and quantization rule $\gamma$, the problem of determining \begin{align}
\phi(f)=\inf_\gamma \sup_{\theta \in (-1,1)} \text{CRB}(\theta,\gamma,f) \label{def:phi}
\end{align}
has been shown to be extremely difficult and remains unsolved \cite{Chen10},\cite{Venkita07}. We refer to the minimizer of \eqref{def:phi} as the \emph{minimax-CRB} quantizer. Our goal in this paper is to design minimax-CRB quantizers for arbitrary noise densities.

\subsection{Previous work}
The problem of quantizer design for minimax-CRB criterion has been addressed only in terms of some scattered results till now. It is well known that the Threshold Quantizer ($\gamma_T$, see \eqref{qtzr:th:prob}), though widely used in the literature \cite{Rib06},\cite{Venkita07}, is unsuitable for high-SNR situations because the maximum value of CRB, typically occurring at boundaries, may exponentially increase with decreasing variance $\sigma^2$ \cite{Papado01}. This problem is often addressed by adding some additional noise (dithering) to the observation prior to threshold-quantization. We refer to this as \emph{Dithering Quantizer} ($\gamma_D$). Dithering is often necessary only in the high-SNR situations, when the noise variance is below a critical magnitude, say $\sigma_\mathcal{F}^2$. The critical variance depends on the shape of the noise pdf and is determined by \cite{Papado01}
\begin{align}
    \sigma_\mathcal{F} = \mbox{ arg } \inf_{\sigma} \phi(\gamma_T,f(\sigma^2)). \label{def:critical:sigma}
\end{align}
By design, the Dithering Quantizer has the limitation that the maximum-CRB actually flattens out (does not decrease) below the critical variance, e.g., for Gaussian noise it was shown in \cite{Papado01} that $\sigma_\mathcal{F} \approx 2/\pi$.
%

\emph{Zero-noise performance limit and Sine Quantizer ($\gamma_0$): } The performance limit of $\phi(f)$ for the noiseless situation, i.e., when $f(w)=\delta(w)$, was derived in \cite{Chen10}. For such a scenario, the optimum minimax-CRB quantizer and the corresponding performance were shown to be
\vspace{-0.05in}
\begin{align}
\gamma_0(x)&\triangleq \left\{
\begin{array}{ll}
1, & x> 1 \\
\frac{1}{2}\left(1+\sin \frac{\pi x}{2}\right), & x\in[-1,1]\\
0, & x< -1
\end{array} \right. , \mbox{ and } \label{res:noiseless:gamma} \\
\phi_0&\triangleq \frac{4}{\pi^2}\approx 0.4. \label{res:noiseless:phi}
\end{align}

It must be noted here that analogous performance limits for finite-variance noise densities are extremely challenging and their derivation remains an open problem. While the quantizer given by \eqref{res:noiseless:gamma} is insightful, it has limited applicability due to two reasons, (1) the noiseless scenario can only approximate high-SNR cases and (2) even for high-SNR cases, $\gamma_0$ may be far from satisfying the minimax property, as we shall show later in this paper.

In this paper, we make some significant contributions towards the study of minimax-CRB quantizer design. We define antisymmetric quantizers and restrict our attention within that class. We determine certain conditions under which the shape of the optimal quantizer is greatly simplified, thereby enabling efficient implementation. We then identify a class of noise distributions for which the Threshold Quantizer is optimal. Lastly for other noise distributions, aided by some theoretical insights, we propose a class of piecewise-linear quantizers and formulate the quantizer design problem as one of numerical minimax optimization. The resulting quantizer is shown to perform significantly better compared to all three existing quantizers - namely the Threshold, Dithering and Sine quantizers.


\section{Main Results} \label{sec:main}
Before presenting the results, we provide some definitions that will be needed for subsequent discussion.

\begin{def:admissible} \label{def:admissible:lbl}
A quantizer $\gamma(x)$ is \emph{admissible} if the resulting conditional probability distribution $g(\theta)$ is monotonically increasing in $\theta\in (-1,1)$.
\end{def:admissible}

The monotonic property is desirable since it ensures that $g^{-1}(\cdot)$ exists so the ML-estimator  \eqref{theta:ml} is well-defined. The increasing property is without loss of generality, since, corresponding to every $\gamma(x)$, there is another valid quantizer $\overline{\gamma}(x)\triangleq 1-\gamma(x)$ such that $\overline{g}(\theta)\triangleq \int_{-\infty}^{\infty} \overline{\gamma}(x) f(x-\theta) \ud x=1-g(\theta)$. This reverses the increasing/decreasing property and yet has the same maximum-CRB, since by \eqref{Itheta:g:g1},  $\text{CRB}(\theta,\gamma,f)=\text{CRB}(\theta,\overline{\gamma},f)$. Hence it is sufficient that, in pursuit of a minimax-CRB quantizer, we restrict our attention to admissible quantizers. Alternatively, throughout the rest of the paper, any reference to a minimax-CRB quantizer will imply that it is admissible.

\begin{def:antisymmetric}
A quantizer $\gamma(x)$ is \emph{antisymmetric} if
\begin{align}
\gamma(x)+\gamma(-x)=1, \qquad \forall x. \label{prop:antisymmetric}
\end{align}
\end{def:antisymmetric}
It may be noted here that traditional quantizers like the Threshold, Dithering, and Sine quantizers are antisymmetric. It is easy to see that antisymmetric property of $\gamma(x)$ together with the assumption of symmetric noise pdf $f(w)$ implies that $g(\theta)$ is also antisymmetric, i.e., $g(\theta)=1-g(-\theta)$. This further means that,
\begin{align}
\text{CRB}(\theta,\gamma;f)=\text{CRB}(-\theta,\gamma;f), \label{crb:theta:symm}
\end{align}
which imply that we can reduce the interval of interest in \eqref{def:phi:gamma} by a factor of half, i.e., either $\theta\in (-1,0]$ or $\theta\in [0,1)$ is sufficient for analysis.

We note here that for an antisymmetric quantizer with symmetric noise pdf,  $g(\theta)$ can be simplified as,
\begin{align}
g(\theta)&=F(\theta) + \int_{-\infty}^{0} \gamma(x) \xi(\theta,x) \ud x, \mbox{ where } \label{g:asym:form:inf} \\
\xi(\theta,x)&\triangleq f(x-\theta)-f(x+\theta), \label{def:xi}
\end{align}
and $F(\theta)\triangleq \int_{-\infty}^{\theta} f(w)\ud w$ is the distribution function.

\begin{def:dominant}
We call a quantizer $\gamma_1(x)$ \emph{dominant} over another quantizer $\gamma_2(x)$ if
\begin{align}
\text{CRB}(\theta,\gamma_1;f)\le \text{CRB}(\theta,\gamma_2;f), \quad \forall \theta\in(-1,1).
\end{align}
\end{def:dominant}
Clearly, a dominant quantizer is better in terms of performance, since it ensures a lesser maximum-CRB, i.e., $\phi(\gamma_1,f)\le \phi(\gamma_2,f)$. As a passing remark, it may be pointed here that the reverse is not necessarily true, i.e., lesser maximum-CRB does not necessarily imply dominance.

\begin{def:unimodal}  \label{def:unimodal:lbl}
A probability density function $f(w)$ is \emph{unimodal} if it has only one maxima (at $w=w_0$, say), i.e., $f'(w)> 0,\text{ for } w\in (-\infty,w_0)$ and $f'(w)< 0, \text{ for } w\in (w_0,\infty)$. For example, commonly used Gaussian and Laplacian noise densities are unimodal.
\end{def:unimodal}

In certain cases, the support of a minimax-CRB quantizer can be highly restricted. Lemma \ref{prop:unit:lbl} lays out such a scenario.

\begin{prop:unit} \emph{(Restricting the domain:)} \label{prop:unit:lbl}
Assume the noise density $f(w)$ to be zero-mean, symmetric and unimodal. Then an antisymmetric minimax-CRB quantizer is at most \emph{unit-support} in the negative semi-axis, i.e.,
\vspace{-0.05in}
\begin{align}
\gamma(x)=0 \mbox{ for } x < -1. \label{prop:unitsupport}
\end{align}
\vspace{-0.05in}
\end{prop:unit}

To establish Lemma \ref{prop:unit:lbl} we show that, for any antisymmetric $\gamma(x)$, there exists a unit-support quantizer $\widetilde\gamma(x)$ (namely, the trivially truncated quantizer),
\vspace{-0.05in}
\begin{align}
\widetilde\gamma(x)\triangleq \left\{\begin{array}{ll}
1        , & x>1 \\
\gamma(x), & x\in[-1,1] \\
0        , & x<-1
\end{array} \right. ,\label{gammatr}
\vspace{-0.05in}
\end{align}
that is both antisymmetric and dominant over $\gamma(x)$. The full proof is provided in Appendix \ref{app:prop:unit}.

The unit-support property helps make the quantizer structure simpler, which will be key in a subsequent theoretical result as well as our numerical design in Section \ref{sec:aupl}. We note here that for an antisymmetric unit-support quantizer with symmetric noise pdf,  $g(\theta)$ and $g'(\theta)$ can be simplified as,
\begin{align}
g(\theta)&=F(\theta) + \int_{-1}^{0} \gamma(x) \xi(\theta,x) \ud x, \mbox{ and } \label{g:asym:form} \\
g'(\theta)&=f(\theta) + \frac{\ud }{\ud \theta}\left\{\int_{-1}^{0} \gamma(x) \xi(\theta,x) \ud x\right\}. \label{dg:asym:form}
\end{align}

In certain cases, the Threshold Quantizer ($\gamma_T$) is also the minimax-CRB quantizer, an example of which is provided in Theorem \ref{prop:threshold:lbl}.

\begin{prop:threshold} \emph{(Optimality of Threshold Quantizer:)} \label{prop:threshold:lbl}
Assume the noise density $f(w)$ to be zero-mean, symmetric, unimodal and such that
\begin{align}
f'(w-z)+f'(w+z)\le 0, \text{ for } w\in[0,1],z\in[0,1]. \label{prop:threhold:cond}
\end{align}
Then, the Threshold Quantizer is dominant over all possible antisymmetric quantizers.
\end{prop:threshold}

The proof of Theorem \ref{prop:threshold:lbl} is given in Appendix \ref{app:threshold}. This is an important result, since condition \eqref{prop:threhold:cond} is satisfied for a wide family of noise densities, including the following example.

\emph{Example \ref{prop:threshold:lbl}.1: }  \label{example:threshold:lbl} \emph{Gaussian density: }
For Gaussian density with variance $\sigma^2$, it is easy to see that condition \eqref{prop:threhold:cond} holds for $\sigma^2\ge 1$ (derivation in Appendix \ref{app:gaussian:sig}). Therefore, for Gaussian noise with variance $\sigma^2\ge 1$, no probabilistic quantizer (within the antisymmetric class) can decrease the maximum-CRB beyond the Threshold Quantizer.

We end this section by pointing out a deficiency of the Sine Quantizer that we alluded to in the introduction. We show that the CRB at the boundaries ($\theta=\pm 1$) for vanishingly small variance ($\sigma^2\rightarrow 0$) is more than twice of that predicted for the noiseless case. The exact degree of sub-optimality depends on the shape of the noise density and is summarized in Proposition \ref{prop:limit:lbl} below.

\begin{prop:limit} \emph{(High-SNR sub-optimality of Sine Quantizer.)} \label{prop:limit:lbl}
Let $f(w;\sigma^2)$ denote a family of zero-mean, symmetric noise densities with $\sigma^2$ signifying the variance. Assume that the moment condition $\sigma^{-4}\int_{-\infty}^{\infty}w^4f(w;\sigma^2)\ud w < \infty$ is satisfied. Then,
\begin{align}
\lim_{\sigma^2\rightarrow 0} \phi(\gamma_0,f(w;\sigma^2)) &\ge \lim_{\sigma^2\rightarrow 0} \text{CRB}(\pm 1,\gamma_0;f(w;\sigma^2))  \label{prop:limit:boundary} \\
&= \frac{4}{\pi^2}\frac{1}{2\mu_1^2} \label{prop:limit:mu1} \\
&> \frac{8}{\pi^2} \label{prop:limit:final} ,
\end{align}
where $\mu_1$ is the normalized one-sided mean, $\mu_1\triangleq\sigma^{-1}\int_{0}^{\infty}w f(w;\sigma^2) \ud w$.
\end{prop:limit}

The proof of Proposition \ref{prop:limit:lbl} is provided in Appendix \ref{app:prop:limit}. The bound $\frac{8}{\pi^2}$ in Proposition \ref{prop:limit:lbl} can be compared directly with the theoretical limit $\frac{4}{\pi^2}$ \eqref{res:noiseless:phi} to note that it is twice as large. For illustration, the specific limit in \eqref{prop:limit:mu1} for Gaussian and Laplacian pdf is tabulated in Table \ref{tbl:limit} (derivation in Appendix \ref{app:prop:limit:eg}). We will further substantiate these results numerically in Section \ref{sec:results}. In terms of a low-noise sensing application with a pre-specified allowable distortion, Proposition \ref{prop:limit:lbl} quantifies the scope of improvement over Sine Quantizer - by a \emph{judicious} design of quantizer (detailed subsequently in Section \ref{sec:aupl}), we can potentially reduce the required number of sensors to half.

\begin{table}
\begin{center}
\begin{tabular}{|c|c|c|c|}
  \hline
  Laplacian & Gaussian & Noiseless case \\
  \hline
  $16/\pi^2\approx 1.62$ & $4/\pi\approx 1.27$ & $4/\pi^2\approx 0.41$ \\
  \hline
\end{tabular}
\end{center}
\caption{Lower bound on maximum-CRB using Sine Quantizers.} \label{tbl:limit}
\vspace{-0.35in}
\end{table}

Proposition \ref{prop:limit:lbl} highlights the sub-optimality of the Sine Quantizer, which necessitates an alternative quantizer design in the high-SNR regime. Even in the moderate-SNR regime, in the absence of concrete analytical results for finite variance scenarios, it is not clear how one should design efficient minimax-CRB quantizers. In the following section, we describe a quantizer design method through direct numerical optimization.

\section{Antisymmetric minimax-CRB quantizer as an optimization problem } \label{sec:aupl}
A general probabilistic quantizer $\gamma(x)$ is any function that maps $(-\infty,\infty)\rightarrow [0,1]$. But numerical search within such a functional space is extremely difficult and hence we make some additional assumptions.

First, the proposed quantizer $\gamma_P(x)$ is assumed to be antisymmetric, and the noise density is assumed to be symmetric. From Lemma \ref{prop:unit:lbl}, this also means that it is unit-support. To further simplify the structure, we assume that $\gamma_P(x)$ is piecewise linear. Hence, we divide the support interval $[-1,0]$ into several equally spaced intervals. We choose the observation grid-size $\Delta_x$ or the number of grid intervals $K$ so that $K\Delta_x=1$. Define $a_0\equiv 0$ and for $k=1,2,\ldots,K$, the following,
\vspace{-0.07in}
\begin{align}
\begin{split}
\mathcal{D}_k &\triangleq  [x_{k-1},x_k], \mbox{ where } x_i=-(K-i)\Delta_x\\
\gamma_P(x)&=a_{k-1}+m_k (x-x_{k-1}), \mbox{ for } x\in \mathcal{D}_k, \mbox{ and } \\
a_{k}&=a_{k-1}+m_k \Delta_x,
\end{split} \label{qtzr:aupl}
\end{align}
where $m_1,m_2,\ldots,m_K$ are the slopes that need to be chosen.

\emph{Notation: } Henceforth, we will refer to the quantizer $\gamma_P(x)$ as the
\emph{Antisymmetric Unit-support Piecewise-Linear (AUPL)} quantizer. The AUPL quantizer is entirely specified in terms of the slope vector $\bo m$.

\emph{Objective Function: }
We characterize the objective function in terms of $\bo m$. For the piecewise linear quantizer $\gamma_P$, the expressions \eqref{g:asym:form} and \eqref{dg:asym:form} reduce to linear functions of $\bo m$, i.e.,
\begin{equation}
\begin{split}
g(\theta)&=[\bo a(\theta)]^T \bo m+F(\theta), \mbox{ and } \\
g'(\theta)&=[\bo c(\theta)]^T \bo m+f(\theta), \mbox{ where } \\
\bo a(\theta)&= J\bo q(\theta)+\bo r(\theta), \\
\bo c(\theta)&= J\bo q'(\theta)+\bo r'(\theta), \\
J&\triangleq \begin{bmatrix}
K & 1 & \cdots & 1 \\
0 & K-1 & \cdots & 1 \\
\vdots & \ddots & \ddots & \vdots \\
0 & 0 & \cdots & 1 \\
\end{bmatrix}, \\
\left[\bo q(\theta)\right]_k & = \Delta_x \int_{\mathcal{D}_k} \xi(\theta,x) \ud x, \mbox{ and } \\
\left[\bo r(\theta)\right]_k & = \int_{\mathcal{D}_k} x \xi(\theta,x) \ud x, \quad k=1,2,\ldots,K.
\end{split}
\end{equation}

Next we discretize the parameter set. We note that the region of interest is only $\theta\in [-1,0]$, with the other half taken care of through symmetry. We choose the parameter grid size $\Delta_\theta$ or the number of grid partitions $L$ so that $L\Delta_\theta=1$. Let the discrete points be
\begin{align}
\theta_l=-\frac{l}{L}, \mbox{ for } l=0,1,\ldots,L
\end{align}
Next, the maximum-CRB due to quantizer $\gamma_P$ (see \eqref{def:phi:gamma}) is approximated as
\begin{align}
\phi(\bo m,f)=\max_{l} \frac{(\bo a_l^T\bo m+F_l) (1-\bo a_l^T\bo m-F_l)}{(\bo c_l^T\bo m+f_l)^2}, \label{crb:aupl:m}
\end{align}
where $\bo a_l\triangleq \bo a(\theta_l)$, $\bo c_l\triangleq \bo c(\theta_l)$, $F_l\triangleq F(\theta_l)$ and $ f_l\triangleq f(\theta_l)$. In Equation \eqref{crb:aupl:m}, $\phi(\cdot,f)$ is our objective function with $\bo m$ as the variable.

\emph{Constraints: } We identify two constraints. Firstly, the slopes $m_k$ must be chosen so that the probability values for all observations $x$ satisfy $\gamma_P(x) \in [0,1]$. Since $\gamma_P$ is piecewise linear, this is ensured by placing inequality constraints at the boundary points. From \eqref{qtzr:aupl}, we obtain $\gamma_P(x_k)=\Delta_x \sum_{j=1}^k m_j=\frac{1}{K} \sum_{j=1}^k m_j$. Hence the probability constraint at point $x_k$ can be expressed as $0\le \sum_{j=1}^k m_j \le K$, for $k=1,2,\ldots,K$. Secondly, from the antisymmetric property, assuming that $\gamma_P(x)$ is continuous at $0$, we have $\gamma_P(0+)=\gamma_P(0-)=1/2$, and hence we need to ensure that  $\gamma_P(0)=\gamma_P(x_K)=1/2$, or equivalently, $\sum_{j=1}^K m_j=K/2$.

\emph{Optimization Problem: }
Finally, the minimax-CRB quantizer $\phi(f)$ defined by \eqref{def:phi} can be obtained as a solution to the following optimization problem in $\mathbb{R}^K$,
\begin{equation}
\begin{split}
\minimize_{\bo m} \quad & \phi(\bo m,f), \\
\mbox{s.t.} \qquad &
\begin{bmatrix}
-\bo L \\
\bo L
\end{bmatrix} \bo m \le
\begin{bmatrix}
\bo 0 \\
K \bo i_K
\end{bmatrix}, \mbox{ and } \\
&(\bo i_K)^T \bo m=K/2, \mbox{ where }
\end{split} \label{prob:max:phi}
\end{equation}
\begin{align}
\bo L&\triangleq \begin{bmatrix}
1 & 0 & \cdots & 0 \\
1 & 1 & \ddots & 0 \\
\vdots & \vdots & \ddots & \vdots \\
1 & 1 & \cdots & 1 \\
\end{bmatrix}, \mbox{ and } \bo i_K\triangleq  \begin{bmatrix}
1 \\
1 \\
\vdots \\
1
\end{bmatrix}.
\end{align}

\emph{Implementation Notes: } It may be noted that $\bo a_l,\bo c_l,F_l,f_l$ in $\phi(\bo m,f)$ (see \eqref{crb:aupl:m}) are all constants and may be pre-computed before running the optimizer. Also, once the noise pdf is known, the optimum quantizer $\gamma_P $ can be computed offline and programmed into the sensor nodes. Choice of $K$ and $L$ essentially provides a tradeoff between discretization artifacts and numerical complexity. From numerical experiments, $K=L \approx \lceil 10/\sigma \rceil$ was found to yield sufficiently convergent results. The problem given by \eqref{prob:max:phi} is not known to be convex (to the best of authors' knowledge) and hence we require multiple and good starting points $\bo m_0$ to obtain a satisfactory solution. In our implementation, we have chosen two starting points for $\bo m_0$, namely the closest AUPL counterparts for  the Threshold and Sine quantizers. We have used the MATLAB function FMINCON for optimization.


\section{Illustrative Examples} \label{sec:results}
We illustrate some of the key ideas in this paper through numerical results.

\begin{figure}[t]
\begin{center}
    \includegraphics[width=\figsizefull \columnwidth]{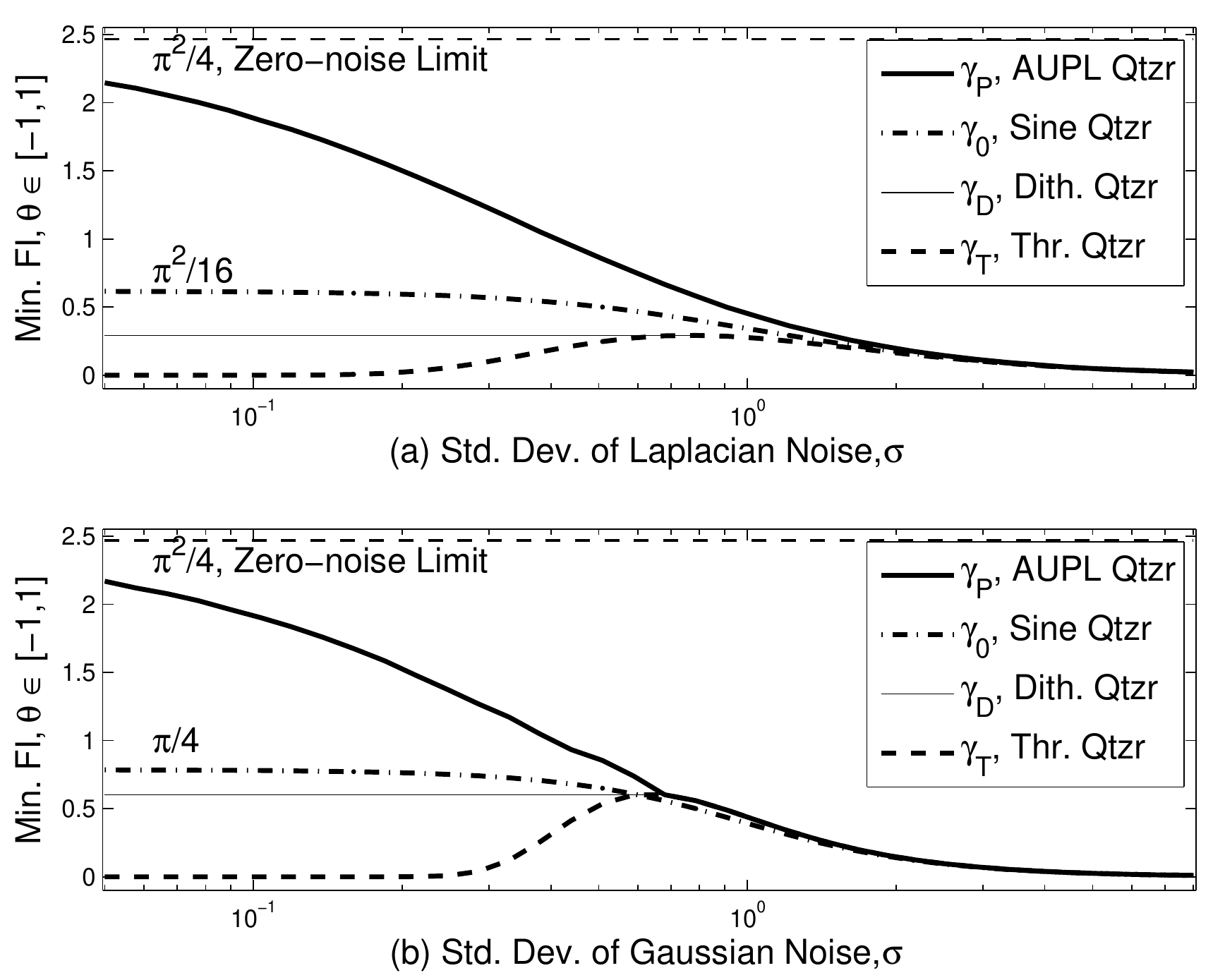}
  \caption{Minimum Fisher information ($\phi(\gamma,f)^{-1}$) for Threshold, Sine, Dithering and AUPL quantizers for (a) Laplacian and (b) Gaussian noise.}
  \label{fig:comp:minFI:LN}
  \end{center}
\vspace{-0.2in}
\end{figure}

\emph{Sub-optimality of Sine Quantizer: }
In Proposition \ref{prop:limit:lbl}, we showed that the Sine Quantizer given by \eqref{res:noiseless:gamma}, though optimum for zero-noise, is significantly sub-optimal when $\sigma$ is small but finite (high-SNR). The results displayed in Figure \ref{fig:comp:minFI:LN} illustrate this phenomena. We display the minimum Fisher Information (inverse of Crame\'r-Rao bound) of the Sine Quantizer. As illustrative noise pdf-s, we consider Gaussian and Laplacian densities over a wide range of variance ($0.05\le \sigma \le 8$). The dotted line showing $\phi_0^{-1}=\pi^2/4$ is the zero-noise limit. The dash-dotted lines corresponding to $\pi^2/16$ and $\pi/4$, which are significantly less than the zero-noise limit, denote the performance of the Sine Quantizer. These results are consistent with the limits described in Table I.

\emph{Performance of AUPL quantizer: }
In Figure \ref{fig:comp:minFI:LN}, we have also compared the AUPL quantizer $\gamma_P$ with the Threshold $\gamma_T$, Dithering $\gamma_D$ and Sine $\gamma_0$ quantizers. The critical standard deviation for Dithering Quantizer correspond to the maxima of the performance of $\gamma_T$ (recall \eqref{def:critical:sigma}). In Figure \ref{fig:comp:minFI:LN}, $\gamma_D$ corresponds to the unbroken horizontal lines connected to the maxima of $\gamma_T$ performance curves. These critical variances are seen to approximately $\sigma_\mathcal{L} \approx 0.79$ and $\sigma_\mathcal{N} \approx 0.63$ for Laplacian and Gaussian noise respectively. We observe that the AUPL quantizer performs better than all three existing quantizers, and considerably so in the moderate to high-SNR regime.

\emph{Minimax-optimality of Threshold Quantizer: }
We showed in Example \ref{prop:threshold:lbl}.1 that for Gaussian density with $\sigma \ge 1$, the Threshold Quantizer is also the antisymmetric minimax-CRB quantizer. We verify in Figure \ref{fig:comp:minFI:LN}-(b) that the performance curves for AUPL and Threshold quantizers coincide for $\sigma\ge 1$. In fact, they seem to coincide somewhat earlier, around $\sigma\ge 0.7$. This is because dominance (see Theorem \ref{prop:threshold:lbl}) is only a sufficient condition for minimax-CRB superiority. It may also be noted that no such coincidence is observed for the Laplacian case (Figure \ref{fig:comp:minFI:LN}-(a)). Since the Laplacian density is not differentiable at the origin, Theorem \ref{prop:threshold:lbl} does not apply in this case.

\begin{figure}[t]
\begin{center}
    \includegraphics[width=\figsizefull \columnwidth]{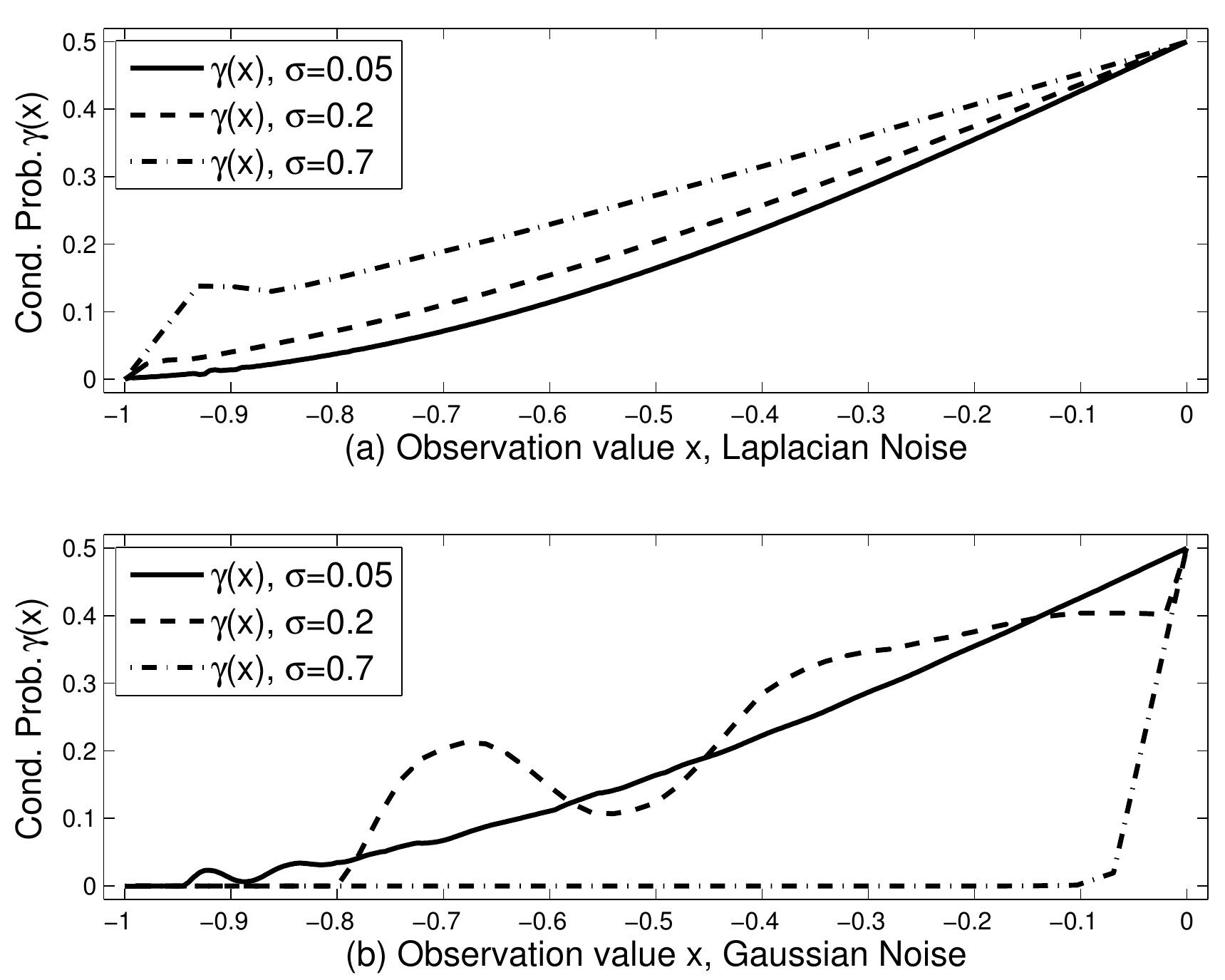}
  \caption{Designed probability $\gamma_P(x)=P(Y=\mathcal S_1|x)$ for AUPL quantizer.}
  \label{fig:shape:aupl:gam}
  \end{center}
\vspace{-0.2in}  
\end{figure}

\begin{figure}[t]
\begin{center}
    \includegraphics[width=\figsizefull \columnwidth]{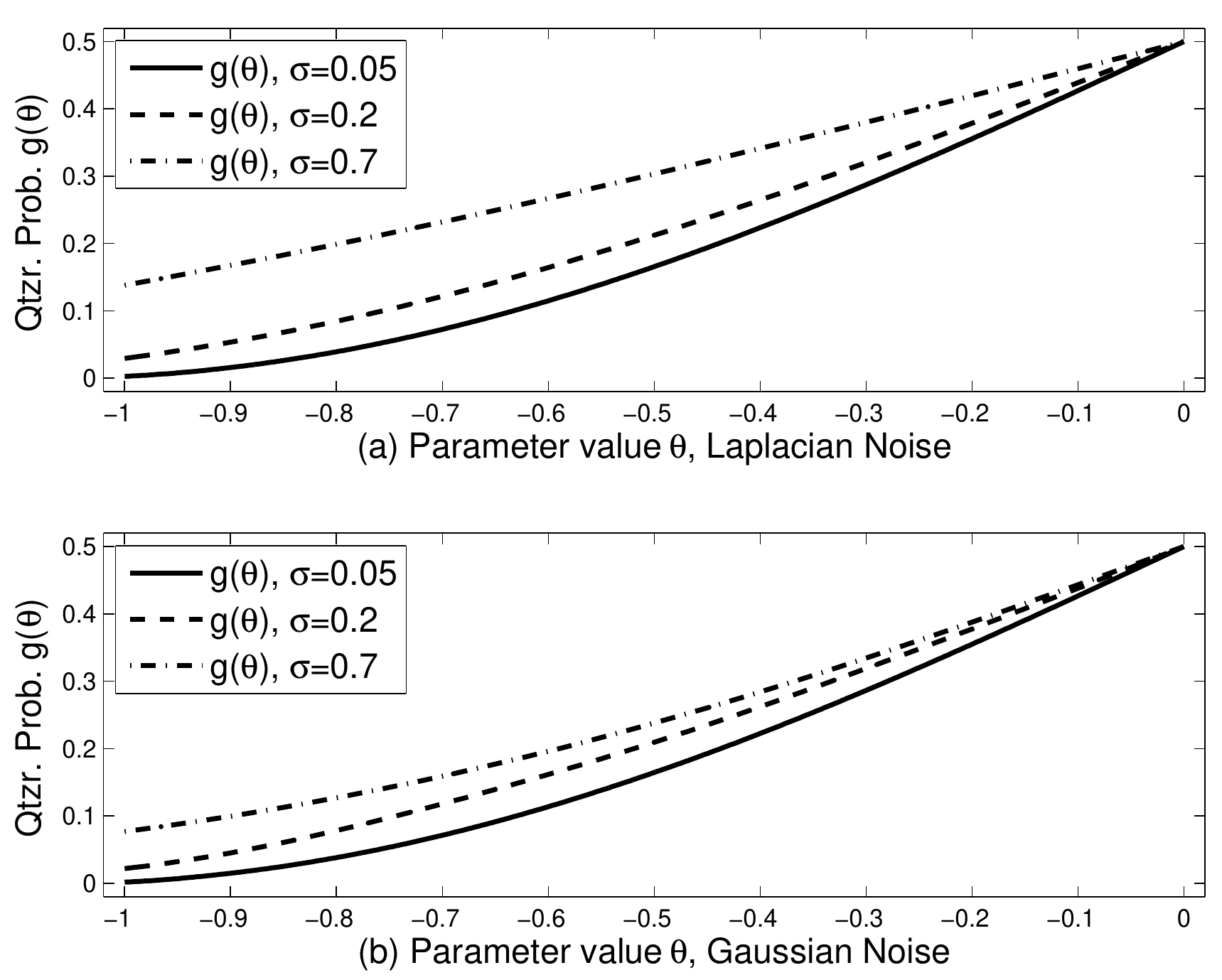}
  \caption{Overall probability $g(\theta)=\mathbb E_W(\gamma_P(\theta+W))$ for AUPL quantizer.}
  \label{fig:shape:aupl:g}
  \end{center}
\vspace{-0.3in}  
\end{figure}

\emph{Shape of AUPL quantizer: }
We display the shape of AUPL quantizer $\gamma_P(x)$ and corresponding $g(\theta)$ for various noise pdf-s in Figures \ref{fig:shape:aupl:gam} and \ref{fig:shape:aupl:g} respectively. We consider Laplacian and Gaussian pdf-s for small ($\sigma=0.05$), medium ($\sigma=0.2$) and large ($\sigma=0.7$) variances. We note that for Gaussian noise, the AUPL quantizer displays a \emph{damped oscillating} behavior, where the bumps get smaller but more in number, with decreasing variance. In the limit of small $\sigma$, the AUPL quantizer is seen to approach the shape of the Sine Quantizer, though not exactly. In the limit of large $\sigma$, for the Gaussian case, the AUPL quantizer is seen to approach the shape of the Threshold Quantizer. Figure \ref{fig:shape:aupl:gam} also shows that $\gamma_P$ need not be monotonic. This is in contrast with commonly used Threshold, Dithering and Sine quantizers, all of which are monotonic. The AUPL quantizer relaxes this assumption and allows for non-monotone functions. The overall quantizer probability $g(\theta)$, however, has to be monotonically increasing in $\theta\in(-1,1)$ to satisfy the admissibility property (see Definition \ref{def:admissible:lbl}). This can be verified in Figure \ref{fig:shape:aupl:g}.

\section{Conclusion} \label{sec:conc}
In this paper, we studied the design of identical binary quantizers for distributed estimation using minimax Crame\'r-Rao lower bound as the performance criterion. Among other theoretical results, we have specified a broad family of distributions for which the Threshold Quantizer is optimal. Aided with some theoretical results, we formulated a numerical optimization problem to obtain the minimax-CRB quantizer within the antisymmetric and piecewise-linear class. We demonstrated the superior performance of the AUPL quantizer for a wide range of noise density functions. Though AUPL quantizers can demonstrably achieve better performance, there are some drawbacks of the same that deserve mention. Firstly, AUPL quantizers are more difficult to implement because of the numerical complexity involved in the design process. Traditional quantizers like Sine, Threshold and Dithering quantizers are relatively simpler to design. Secondly, since AUPL quantizer is tailored to a particular noise density, it may not be suitable for applications where the ambient noise changes frequently. Lastly, the AUPL quantizer in Section \ref{sec:aupl} was derived under the assumption of noiseless channels. Extension of AUPL quantizer to noisy communication channels merit further investigation.



\appendices
\section{Proof of Lemma \ref{prop:unit:lbl}} \label{app:prop:unit}
Let $\widetilde g(\theta)$ be the conditional probability corresponding to $\widetilde\gamma(x)$. We need to show that $\widetilde\gamma(x)$ dominates $\gamma(x)$ for the (half-range) $\theta\in[0,1)$. Refer to the expression of $I(\theta)$ in \eqref{Itheta:g:g1}. It suffices to show that the numerator and denominator terms satisfy, for $\theta\in[0,1)$, the inequalities (N) $(g')^2 \le (\widetilde g')^2$ and (D) $g(1-g) \ge \widetilde g(1-\widetilde g)$. Since admissibility (Definition \ref{def:admissible:lbl}) implies $g'>0$ and $g\ge 1/2$ for $\theta\in[0,1)$, it suffices to show that (N1) $g'\le \widetilde g'$ and (D1) $g\le \widetilde g$. From \eqref{g:asym:form:inf}, we obtain
\begin{align}
g-\widetilde g =\int_{-\infty}^{-1} \gamma(x) (f(x-\theta)-f(x+\theta)) \ud x. \label{g:gtr}
\end{align}
With unimodality of $f(w)$ implying $f(x-\theta)\le f(x) \le f(x+\theta)$, for $x\in(-\infty,-1),\theta\in[0,1)$ and $\gamma(x)$ being positive by definition, (D1) is established from \eqref{g:gtr}. Since $f'(w)> 0$ in $(-\infty,0)$ (see Definition \ref{def:unimodal:lbl}), we can interchange the order of integration and derivative in \eqref{g:gtr}, to obtain,
\begin{align*}
g'-\widetilde g'=-\int_{-\infty}^{-1} \gamma(x) (f'(x-\theta)+f'(x+\theta)) \ud x \le 0,
\end{align*}
thereby establishing (N1).

\section{Proof of Theorem \ref{prop:threshold:lbl}} \label{app:threshold}
Note that conditions for Lemma \ref{prop:unit:lbl} are satisfied, hence it suffices to show that the Threshold Quantizer dominates any admissible antisymmetric unit-support quantizer $\gamma(x)$. Refer to expression of $I(\theta)$ in \eqref{Itheta:g:g1}. It suffices to show that the numerator and denominator terms satisfy, for $\theta\in[0,1)$, the inequalities (N) $(g')^2\le (f)^2$ and (D) $g(1-g)\ge F(1-F)$. Since admissibility (Definition \ref{def:admissible:lbl}) implies $g'>0$ and $g \ge 1/2$ for $\theta\in[0,1)$, it suffices to show that (N1) $g'\le f$ and (D1) $g\le F$. From the definition of $\xi(\theta,x)$ in \eqref{def:xi} and condition \eqref{prop:threhold:cond}, we have for $x\in [-1,0],\theta\in [0,1)$
\begin{align}
\frac{\ud}{\ud \theta}\xi(\theta,x)=-(f'(x-\theta)+f'(x+\theta)) \le 0. \label{cond:1a}
\end{align}
Since $\gamma(x)$ is always positive, Equations \eqref{dg:asym:form} (with interchanged order of integration and differentiation) and \eqref{cond:1a} yield (N1). By integrating Equation \eqref{cond:1a} along $\theta\in[0,\theta_1]$ with the boundary condition $\xi(0,x)=0$ (which is true by definition), we obtain
\begin{align}
\xi(\theta_1,x)&\le 0, \mbox{ for } x\in[-1,0],\theta_1\in[0,1). \label{cond:1b}
\end{align}
Once again, since $\gamma(x)$ is always positive, Equations \eqref{g:asym:form} and \eqref{cond:1b} yield (D1).

\section{Derivation for Example 1.1} \label{app:gaussian:sig}
We will show that \eqref{prop:threhold:cond} holds for Gaussian distribution with $\sigma^2\ge 1$. Since unimodality ensures that $f'(w)< 0$ for $w>0$, it suffices to show that \eqref{prop:threhold:cond} hold in the (restricted) domain $0\le z\le w \le 1$. Noting that $f'(w-z)=-f'(z-w)$ (from symmetric property of $f(w)$) and defining $\alpha \triangleq w/z$, it suffices to establish
\begin{align}
f'(z(1+\alpha))\le f'(z(1-\alpha)) \label{cond:ex1}
\end{align}
for the domain $\alpha\in[0,1]$, $z\in[0,1]$ and $\sigma^2\ge 1$. Substituting $f'(w)=$ $-(\sqrt{2\pi}\sigma^3)^{-1}w\exp{(-w^2/(2\sigma^2))}$ and rearranging terms, condition \eqref{cond:ex1} is equivalent to showing
\begin{align}
\log\frac{1+\alpha}{1-\alpha}\ge \frac{2\alpha z^2}{\sigma^2}. \label{cond:ex2}
\end{align}
The following identity can be ascertained easily for $0 \le \alpha \le 1$
\begin{align}
h(\alpha)\triangleq \log\frac{1+\alpha}{1-\alpha}-2\alpha \ge 0, \label{cond:ex3}
\end{align}
by noting that $h(0)=0$ and $h'(\alpha)\ge 0$. The additional conditions $0 \le z \le 1$ and $\sigma^2 \ge 1$ imply  \eqref{cond:ex2}, thereby completing the derivation.

\section{Proof of Proposition \ref{prop:limit:lbl}} \label{app:prop:limit}
We would prove Equation \eqref{prop:limit:mu1} and \eqref{prop:limit:final}. Starting from \eqref{g:asym:form}, for $\theta=-1$ and small $\sigma$ we proceed from \eqref{g:asym:form} as follows
\begin{align}
\begin{split}
g(-1)&=F(-1) + \int_{-1}^{0} \gamma_0(x) \left( f(x+1)-f(x-1) \right) \ud x \\
 &\stackrel{(a)}{=}F(-1) + \int_{0}^{1} \gamma_0(z-1) \left( f(z)-f(z-2) \right) \ud z \\
 &\stackrel{(b)}{=}\int_{0}^{\infty} \gamma_0(z-1) f(z) \ud z +\mathcal O(\sigma^4)\\
 &\stackrel{(c)}{=}\int_{0}^{\infty} \left(\frac{\pi^2}{16}z^2+\mathcal{O}(z^4)\right) f(z) \ud z +\mathcal O(\sigma^4) \\
 &\stackrel{(d)}{=}\frac{\pi^2\sigma^2}{32}+\mathcal O(\sigma^4),
\end{split} \label{g:m1:O}
\end{align}
where (a) is due to change in variables, (b) is due to bounding three distinct terms. First, $F(-1)=\int_{-\infty}^{-1} f(z) \ud z$ can be bounded to $\mathcal O(\sigma^4)$\footnote{Recall the Landau or "big O" notation: a function $f$ is asymptotically bounded above by $g$, written $f(n)=\mathcal{O}(g(n))$, if there exist constants $N>0$ and $c>0$ such that $f(n)\le c g(n)$ for all $n>N$.} assuming that the normalized fourth-moment is bounded and applying an inequality (precisely, no. $26.1.41$) in \cite{Abramowitz64}. The other two terms are themselves bounded by $F(-1)$, i.e., $\int_1^{\infty} \gamma_0(z-1) f(z) \ud z<\int_1^{\infty} f(z) \ud z=F(-1)$ and $\int_0^1 \gamma_0(z-1) f(z-2) \ud z<\int_0^1 f(z-2) \ud z<F(-1)$ and hence are of the order $\mathcal O(\sigma^4)$. (c) follows from the Taylor-series expansion of \eqref{res:noiseless:gamma} in the vicinity of $z=0$,  \[\gamma_0(z-1)=\frac{1}{2}\left[1-\cos\left(\frac{\pi z}{2}\right)\right]=\frac{\pi^2z^2}{16}+\mathcal{O}(z^4),\]
and (d) follows from the partial moment relation $\int_0^\infty z^n f(z) \ud z=\mathcal O(\sigma^n)$. Similarly, for $g'(\theta)$ in \eqref{dg:asym:form},
\begin{align}
\begin{split}
g'(-1)&=-\frac{\pi^2}{16} \int_{0}^{\infty}z^2 f'(z) \ud z +\mathcal O(\sigma^3) \\
 &\stackrel{(a)}{=}\frac{\pi^2\sigma \mu_1}{8} +\mathcal O(\sigma^3),
\end{split} \label{dg:m1:O}
\end{align}
where (a) follows from integration by parts and the fact that $z^2f(z)|_0^\infty=0$.
Applying \eqref{g:m1:O} and \eqref{dg:m1:O} in \eqref{Itheta:g:g1} we obtain \eqref{prop:limit:mu1}.

The normalized one-sided mean $\mu_1$ depends on the shape of the noise density and the inequality in \eqref{prop:limit:final} is due to the fact that $\mu_1< 1/2$ for any zero-mean, symmetric noise density $f(w)$. Consider the function
\begin{align}
f_0(w)&=\left\{\begin{array}{ll}
2f(w) & w\ge 0 \\
0 & w<0
\end{array} \right.,
\end{align}
which is also a density function since $\int_{-\infty}^{\infty}f_0(w) \ud w=1$ and hence must have a positive variance. Thus $\text{Var}_{f_0}(w)=\mathbb E_{f_0}(w^2) -(\mathbb E_{f_0}(w))^2=\sigma^2(1-4\mu_1^2)>0$. Hence $\mu_1<1/2$.

\section{Derivation of limits in Table \ref{tbl:limit}} \label{app:prop:limit:eg}
To derive the limits in Table \ref{tbl:limit}, we consider the \emph{generalized Gaussian} density \cite{Nadarajah05}, specified in terms of the shape parameter $\beta$ and variance $\sigma^2$ as
$f(w;\beta,\sigma^2)=$$\frac{\beta}{2\alpha\Gamma(1/\beta)} \exp\left(-\left(|w|/\alpha\right)^\beta\right)$,
where $\alpha$ is related to variance by $\alpha^2=\sigma^2 \frac{\Gamma(1/\beta)}{\Gamma(3/\beta)}$ and the one-sided mean is $\int_{0}^{\infty} w f(w) \ud w=\frac{\alpha \Gamma(2/\beta)}{2 \Gamma(1/\beta)}$. Here $\Gamma(b) \triangleq  \int_0^\infty  t^{b-1} e^{-t} \ud t$ denotes the Gamma function. Common densities like Laplacian ($\beta=1$) and Gaussian ($\beta=2$) pdf-s are specific examples of this family. From \cite{Nadarajah05}, the normalized fourth-moment is $\sigma^{-4}\int_{-\infty}^{\infty}w^4f(w;\sigma^2)=\frac{\Gamma(5/\beta)}{\Gamma(1/\beta)}$, which is clearly bounded for finite $\beta$. Hence Proposition \ref{prop:limit:lbl} applies, and we have from \eqref{prop:limit:mu1},
\begin{align}
    \lim_{\sigma^2\rightarrow 0} \text{CRB}(\pm 1,\gamma_0;f(w;\beta,\sigma^2)) &= \frac{8}{\pi^2} \frac{\Gamma(1/\beta)\Gamma(3/\beta)}{(\Gamma(2/\beta))^2}. \label{limit:phi:gen:gauss}
\end{align}
Specific instances of this result $\beta=1$ and $\beta=2$ are shown in Table \ref{tbl:limit}. Note that $\Gamma(1)=1$, and $\Gamma(b)=(b-1)\Gamma(b-1)$ for all $b>1$, which simplifies to $\Gamma(b)=(b-1)!$ for integer $b>1$. Furthermore, $\Gamma(1/2)=\sqrt{\pi}$.

\bibliographystyle{IEEEtran}
\bibliography{source-file-ref}

\begin{thebibliography}{10}
\providecommand{\url}[1]{#1}
\csname url@samestyle\endcsname
\providecommand{\newblock}{\relax}
\providecommand{\bibinfo}[2]{#2}
\providecommand{\BIBentrySTDinterwordspacing}{\spaceskip=0pt\relax}
\providecommand{\BIBentryALTinterwordstretchfactor}{4}
\providecommand{\BIBentryALTinterwordspacing}{\spaceskip=\fontdimen2\font plus
\BIBentryALTinterwordstretchfactor\fontdimen3\font minus
  \fontdimen4\font\relax}
\providecommand{\BIBforeignlanguage}[2]{{%
\expandafter\ifx\csname l@#1\endcsname\relax
\typeout{** WARNING: IEEEtran.bst: No hyphenation pattern has been}%
\typeout{** loaded for the language `#1'. Using the pattern for}%
\typeout{** the default language instead.}%
\else
\language=\csname l@#1\endcsname
\fi
#2}}
\providecommand{\BIBdecl}{\relax}
\BIBdecl

\bibitem{Chen10}
H.~Chen and P.~Varshney, ``Performance limit for distributed estimation systems
  with identical one-bit quantizers,'' \emph{Signal Processing, IEEE
  Transactions on}, vol.~58, no.~1, pp. 466--471, Jan. 2010.

\bibitem{Rib06}
A.~Ribeiro and G.~B. Giannakis, ``Bandwidth-constrained distributed estimation
  for wireless sensor networks-{Part I: Gaussian} case,'' \emph{Signal
  Processing, IEEE Transactions on}, vol.~54, no.~3, pp. 1131--1143, 2006.

\bibitem{Venkita07}
P.~Venkitasubramaniam, L.~Tong, and A.~Swami, ``Quantization for maximin {ARE}
  in distributed estimation,'' \emph{Signal Processing, IEEE Transactions on},
  vol.~55, no.~7, pp. 3596--3605, July 2007.

\bibitem{Gray93}
R.~Gray and J.~Stockham, T.G., ``Dithered quantizers,'' \emph{Information
  Theory, IEEE Transactions on}, vol.~39, no.~3, pp. 805--812, May 1993.

\bibitem{Papado01}
H.~Papadopoulos, G.~Wornell, and A.~Oppenheim, ``Sequential signal encoding
  from noisy measurements using quantizers with dynamic bias control,''
  \emph{Information Theory, IEEE Transactions on}, vol.~47, no.~3, pp.
  978--1002, Mar. 2001.

\bibitem{Luo05}
Z.-Q. Luo, ``Universal decentralized estimation in a bandwidth constrained
  sensor network,'' \emph{Information Theory, IEEE Transactions on}, vol.~51,
  no.~6, pp. 2210--2219, June 2005.

\bibitem{Aysal08}
T.~Aysal and K.~Barner, ``Constrained decentralized estimation over noisy
  channels for sensor networks,'' \emph{Signal Processing, IEEE Transactions
  on}, vol.~56, no.~4, pp. 1398--1410, April 2008.

\bibitem{Wu09}
T.~Wu and Q.~Cheng, ``Distributed estimation over fading channels using one-bit
  quantization,'' \emph{Wireless Communications, IEEE Transactions on}, vol.~8,
  no.~12, pp. 5779--5784, Dec. 2009.

\bibitem{Kay93}
S.~M. Kay, \emph{Fundamentals of Statistical Signal Processing: Estimation
  Theory}.\hskip 1em plus 0.5em minus 0.4em\relax Englewood Cliffs, NJ:
  Prentice Hall, 1993.

\bibitem{Abramowitz64}
M.~Abramowitz and I.~Stegun, \emph{Handbook of Mathematical Functions},
  5th~ed.\hskip 1em plus 0.5em minus 0.4em\relax New York: Dover, 1964.

\bibitem{Nadarajah05}
S.~Nadarajah, ``A generalized normal distribution,'' \emph{Journal of Applied
  Statistics}, vol.~32, no.~7, pp. 685--694, 2005.

\end{thebibliography}

\end{document}